\documentclass[sigconf]{aamas} 
\pdfoutput=1
\usepackage{subcaption}
\usepackage[export]{adjustbox}
\usepackage{balance} % for balancing columns on the final page
\usepackage{multirow}

\usepackage{graphicx}
\usepackage[labelformat=simple]{caption}

\usepackage{balance} % for balancing columns on the final page

\setcopyright{ifaamas}
\acmConference[AAMAS '24]{Proc.\@ of the 23rd International Conference
on Autonomous Agents and Multiagent Systems (AAMAS 2024)}{May 6 -- 10, 2024}
{Auckland, New Zealand}{N.~Alechina, V.~Dignum, M.~Dastani, J.S.~Sichman (eds.)}
\copyrightyear{2024}
\acmYear{2024}
\acmDOI{}
\acmPrice{}
\acmISBN{}

\acmSubmissionID{592}

\title[AAMAS-2024 Formatting Instructions]{Can Poverty Be Reduced by Acting on Discrimination? An Agent-based Model for Policy Making}

\author{Alba Aguilera}
\affiliation{
  \institution{Artificial Intelligence Research Institute, IIIA-CSIC}
  \city{Barcelona}
  \country{Spain}}
\email{aaguilera@iiia.csic.es}

\author{Nieves Montes}
\affiliation{
  \institution{Artificial Intelligence Research Institute, IIIA-CSIC}
  \city{Barcelona}
  \country{Spain}}
\email{nmontes@iiia.csic.es}

\author{Georgina Curto}
\affiliation{
  \institution{
University of Notre Dame}
  \city{Notre Dame}
  \country{United States}}
\email{gcurtore@nd.edu}

\author{Carles Sierra}
\affiliation{
  \institution{Artificial Intelligence Research Institute, IIIA-CSIC}
  \city{Barcelona}
  \country{Spain}}
\email{sierra@iiia.csic.es}

\author{Nardine Osman}
\affiliation{
  \institution{Artificial Intelligence Research Institute, IIIA-CSIC}
  \city{Barcelona}
  \country{Spain}}
\email{nardine@iiia.csic.es}

\begin{abstract}
In the last decades, there has been a deceleration in the rates of poverty reduction, suggesting that traditional redistributive approaches to poverty mitigation could be losing effectiveness, and alternative insights to advance the number one UN Sustainable Development Goal are required. The criminalization of poor people has been denounced by several NGOs, and an increasing number of voices suggest that discrimination against the poor (a phenomenon known as \emph{aporophobia}) could be an impediment to mitigating poverty. In this paper, we present the novel Aporophobia Agent-Based Model (AABM) to provide evidence of the correlation between aporophobia and poverty computationally. We present our use case built with real-world demographic data and poverty-mitigation public policies (either enforced or under parliamentary discussion) for the city of Barcelona. We classify policies as discriminatory or non-discriminatory against the poor, with the support of specialized NGOs, and we observe the results in the AABM in terms of the impact on wealth inequality. The simulation provides evidence of the relationship between aporophobia and the increase of wealth inequality levels, paving the way for a new generation of poverty reduction policies that act on discrimination and tackle poverty as a societal problem (not only a problem of the poor).   
\end{abstract}


\newcommand{\BibTeX}{\rm B\kern-.05em{\sc i\kern-.025em b}\kern-.08em\TeX}

\makeatletter
\gdef\@copyrightpermission{
	\begin{minipage}{0.3\columnwidth}
		\href{https://creativecommons.org/licenses/by/4.0/}{\includegraphics[width=0.90\textwidth]{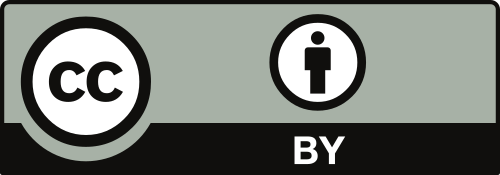}}
	\end{minipage}\hfill
	\begin{minipage}{0.7\columnwidth}
		\href{https://creativecommons.org/licenses/by/4.0/}{This work is licensed under a Creative Commons Attribution International 4.0 License.}
	\end{minipage}
	\vspace{5pt}
}
\makeatother

\begin{document}

\pagestyle{fancy}
\fancyhead{}

\maketitle 

%%%%%%%%%%%%%%%%%%%%%%%%%%%%%%%%%%%%%%%%%%%%%%%%%%%%%%%%%%%%%%%%%%%%%%%%

\section{Introduction}
According to the World Bank \cite{TheWorldBank2023}, over six hundred and fifty million people (10\% of the global population) still live in extreme poverty and COVID-19 has particularly affected the poorest: the number of people living in extreme poverty rose by 11 \% in 2020 \cite{WorldBank2022}. In this context, urgent and innovative measures are required to work towards poverty eradication, the number one UN Sustainable Development Goal. Traditional policies based on the redistribution of wealth could be losing effectiveness, since there has been a deceleration in the poverty reduction rates throughout the last decades \cite{Claudia2018}. Artificial Intelligence tools can provide alternative insights to this global challenge. 

In 2017, the Spanish philosopher Adela Cortina coined the term \textit{aporophobia} to refer to the "rejection, aversion and fear directed toward the poor" \cite{Cortina2022}, which cuts across xenophobia, racism, antisemitism, and other types of discrimination. This concept was officially acknowledged as an aggravating factor for hate crimes in the Spanish legal framework in 2021 \cite{BOE}. The impact of aporophobia in poverty mitigation has been described in the literature: when the poor are rejected and even blamed for their situation, it is more difficult for policymakers to approve poverty-mitigating policies, since these regulations are unpopular \cite{Applebaum2001, Arneson1997, Nunn2009}. However, there is no empirical evidence yet that aporophobia hinders poverty reduction. This paper aims to fill this gap and inform about the effectiveness of poverty-mitigation policies, which consider not only the redistribution of wealth but also the underlying beliefs on the topic of poverty expressed in the regulatory framework.

Commonly, policy measures are thoroughly formulated. However, evidence of the actual societal effects is only obtained once the policies are implemented. Computational simulations can help shed light on the overall effect of public policies and their potential unforeseen effects in a non-invasive way. Agent-Based Modelling (ABM) is a suitable tool for this purpose, as it can be used to model the individual interactions that lead to complex social phenomena, including issues related to law, economy, regulatory design, inequality and poverty \cite{picker, boundarylaw, POLEDNA2023104306, Wicaksono2020}. This article builds on the published architecture of an aporophobia agent-based model for poverty and discrimination policy-making \cite{AMPM}. Here we present the first experimental results and proof of concept.

Our particular approach, the Aporophobia Agent-Based Model (AABM), is a multiagent simulation with autonomous decision-making agents designed to represent citizens, who carry out their daily affairs and are influenced by the legal framework. These individuals have a personal profile based on real-life demographic data. Their decision-making is based on their needs, while the legal norms reflect real-world policies enforced in a specific region (the city of Barcelona has been used as a first use-case scenario). By using this social simulation, we are able to examine the impact that aporophobia, expressed within the legal framework, has on poverty and inequality levels. 

The paper is organized as follows: Section \ref{sec: relworks} underlines the novelty of our work by exploring existing literature on agent-based modelling in policy-making within the applicative domain. In Section \ref{sec: modform}, we explain the methodology used to formulate the model: starting from the fundamental ABM, then adding the autonomous decision-making feature, and finally implementing the regulatory environment. Section \ref{sec: use case} presents the use case of the simulation, i.e. the physical location and the selected regulations implemented. In Section \ref{sec: results}, we discuss the final results, focusing on the individual, collective and overall effect of the norms. Finally, Section \ref{sec: concl} provides insights into the main implications, limitations, and possible avenues for future research.

%\vspace{-0.09cm}

\section{Related Work}
\label{sec: relworks}
Policy-making is a complex process that often leverages various computational tools. Policies, representing legal norms designed by decision-makers, influence behaviour at the individual level. Agent-Based Models (ABMs) inherently adopt a bottom-up, decentralized approach, which provides some advantages in situations where conventional predictive policy-making methods (e.g., Neo-classical Equilibrium Modeling, Traditional Game Theory, System Dynamics, and Serious Gaming) are least effective \cite{enhancingABM, bookintro}. While ABM techniques may not foresee all outcomes of new policies, they serve as valuable tools for uncertainty analysis \cite{bookmultiagent}. %Policies, representing decision-makers' interventions, influence decisions in a top-down manner, though their effect on the environment emerges in a bottom up manner. 

%There are several ABM frameworks described in the literature in the field of poverty-mitigation policy-making \cite{enhancingABM, legaltheorytool}. These mostly aim to inform about poverty alleviation in small community settings, focusing on a specific application instead of taking into consideration the population at large \cite{wossen2015climate, smajgl2013behaviour, dou2020pathways}, to name a few. However, ABMs have not yet tackled the topic of poverty from the perspective of the social values and behaviours \cite{Curto2022}, which are expressed both at a personal and institutional level through the legal system \cite{TerradillosBasoco2020}.   

Existing policy-making frameworks in the ABM literature are highly specialized for specific applications \cite{legaltheorytool}. These mostly aim to inform about poverty alleviation in small community settings, instead of taking into consideration the population at large (e.g. \cite{wossen2015climate, smajgl2013behaviour, dou2020pathways}). However, ABMs have not yet tackled the topic of poverty from the perspective of discriminatory behaviour \cite{Curto2022}, which is expressed both at a personal and institutional level through the legal system \cite{TerradillosBasoco2020}.

%Some focus has been on policy-making for migrational \cite{migrational}, environmental \cite{environmental}, urbanistic \cite{urbananal}, health-related \cite{https://doi.org/10.1111/1475-6773.13916} or agricultural \cite{agricultural} resource management issues, to name a few. In the field of poverty alleviation, ABMs ..... 

Simultaneously, significant efforts are being made to advance the development of Urban Digital Twins (UDT) \cite{UDT}, tailored to serve as realistic representations of urban environments. The agent-based modelling literature is rich on COVID-19 policy-making case studies that require detailed simulation akin to Digital twins \cite{lorig2021agent, alrashed2022covid}. These models must account for an explicit physical and demographic characterization in addition to the social aspects and dynamics of ABMs. Some research about Social Urban Digital Twins (SUDT) \cite{socialdigtwin, socialdigtwin2}, representing the union of both the agent-based modelling and the digital twin community, aims to leverage these computational techniques to enhance the practical utility and real-world impact of simulations. 

This article contributes to the existing ABM literature with a use case that aims to be representative of an urban environment and its underlying social characteristics. Unlike many existing frameworks, our approach is designed for general applicability and scalability. However, the main novelty of the article is the approach to tackling the multidimensional topic of poverty from the perspective of discrimination (aporophobia). The paper also illustrates how we extend the applicability of the needs-based model by Dignum (a COVID-19 case study) \cite{dignum2021social} to other domains, in this case, poverty and discrimination policy-making. 

\section{Model formulation} \label{sec: modform}
To develop our AABM, we have focused on three distinct aspects of the modelling: (i) agent profiles, (ii) agents' autonomous decision-making, and (iii) the environment, which is composed of two elements: the physical and the regulatory \cite{AMPM}. The physical environment represents the virtual space where agents inhabit, interact and take actions, following the classical features of an ABM. The regulatory environment, on the other hand, encompasses the policies, laws and regulations implemented within the system. 

Agents' profiles are a set of selected socio-economic and demographic attributes, extracted from real-world data. The fundamental structure of the simulation is formed by establishing these agents' profiles along with the environment that surrounds them. Agents possess "free will" (autonomous capacity for decision-making), introduced following the needs-based model by Dignum et. al \cite{dignum2020analysing}. The set of legal policies affecting them is specifically selected to target poverty. Our goal is to observe the effect of these policies (input of the model) on the behaviour and levels of wealth of the citizens by examining the final wealth distribution (output of the model). Note that agents' wealth is regarded as their income minus their expenditures. While income inequality (or in this case wealth inequality expressed as income and expenditure) is not a synonym of poverty, several studies have concluded that income inequality affects the pace at which growth enables poverty reduction \cite{Ravallion2004, Aghion1999, Galor2004,Ostry2014}. We acknowledge that poverty is a multidimensional phenomenon \cite{Sen2001}, but in this proof of concept we inform about the effectiveness of poverty-mitigation policies and the impact of aporophobia only in terms of wealth distribution.

The AABM is developed using Python's library Mesa, suitable for building, analyzing and visualizing ABMs \cite{mesa}. The methodology followed involves three basic procedural steps: (1) constructing the Fundamental Model structure, (2) implementing the Needs-Based Model decision-making feature, and (3) integrating the Regulatory Environment. All the project materials can be accessed from the corresponding GitHub public repository. \footnote{https://github.com/albaaguilera/Aporophobia}

\subsection{Fundamental Model} \label{sec: fundabm}
The first step involves building the fundamental structure of our model, which reflects agents living in a city. These agents possess a personal profile based on demographic and socio-economic characteristics instantiated from real-world data. 

Since our AABM is intended for the study of poverty in urban environments, we model a city as a virtual grid with different locations associated with facilities (workplaces, leisure spots, schools, hospitals, etc.). We randomly allocate these facilities within a very simplified $10x10$ grid, which constitutes the "physical environment" of the agents. In future versions of the model, we aim to enhance this geographical representation by using Open Street Map geodata\footnote{Open Street Map web page: \url{https://www.openstreetmap.org/copyright}}.

Agents living in this city are characterized by a unique personal profile, i.e., several associated attributes which can be personal (gender, age and status), economic (wealth, rent and income) and spatial (district and home location).

\subsection{Needs-Based Model} 
\label{sec: needsmodel}
 The second phase of the AABM development involves programming agents with an autonomous decision-making feature based on the needs-based model, first used by Dignum et. al in the ASSOCC model \cite{dignum2021social}. The ASSOCC framework provides a tool to experiment and evaluate possible public health-related interventions and their combined effects in the context of a pandemic \cite{dignum2020analysing}. We will discuss the contextual background behind the model, as well as the most important structures for its functioning: the need satisfaction level function, the expected satisfaction level function and the deliberation function. 
\begin{figure}
    \centering
    \includegraphics[width=0.8\linewidth]{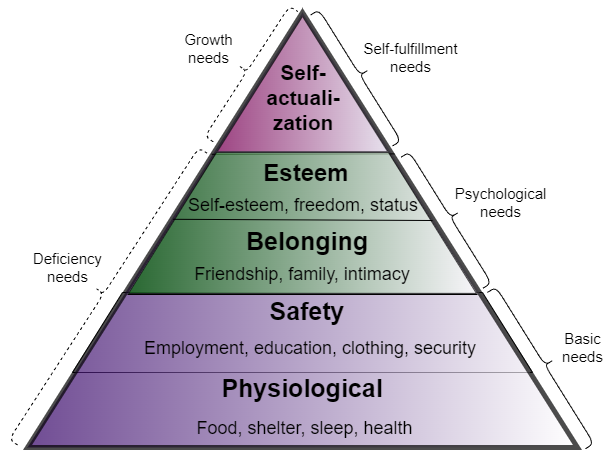}
    \caption{Maslow's hierarchy of needs. Individual needs are divided into categories: Physiological, Safety, Belonging, Esteem and Self-actualization, which in turn can be encapsulated in Basic, Psychological and Self-fulfillment needs, or Deficiency and Growth needs.}
    \Description{Maslow's hierarchy of needs is depicted in a three-main-level pyramid (Basic needs (purple), Psychological needs (green) and Self-fulfillment needs (pink). The individual needs are divided into categories: Psychological, Safety, Belonging, Esteem and Self-actualization, which in turn can be encapsulated in the three main-level already mentioned or in another classification: Deficiency and Growth needs.}
    \label{fig: maslow}
\end{figure}
\subsubsection{Contextual Background}
The Needs-Based model is inspired by Maslow’s hierarchy of needs, a psychological theory of motivation comprising a five-tier model of human needs \cite{maslow}, often depicted as hierarchical levels within a pyramid, as it is illustrated in Figure \ref{fig: maslow}. From the bottom of the hierarchy upwards, the need categories are physiological (including food, shelter, sleep and health), safety (clothing, financial security, employment and education), love and belonging needs (friendship, family and intimacy), esteem (freedom, status and self-esteem), and self-actualization (creativity, morality and acceptance of facts). This five-stage model can be divided into deficiency needs and growth needs. Deficiency needs are concerned with basic survival, essentially considered as a means to an end, whereas growth needs are associated with realizing an individual’s full potential through creative and intellectual activities. Specifically, the needs-based model directs agents to decide on their actions based on how much they expect actions to fulfil their most urgent needs. For instance, if an agent is hungry (\textit{food} need is depleted), their likely choice might be to "\textit{go to grocery store}". 

\subsubsection{Need Satisfaction and Expected Satisfaction}
In our AABM, agents are constrained by their environment, personal profile and needs. These needs and the importance that agents assign to them capture the agents’ societal values and the motives that drive them to action. Agents have a list of available actions determined by their status (e.g. only an employed agent can perform the action “go to workplace”). The status determines how agents can address their most urgent needs. Besides, needs have a certain satisfaction level that requires maintenance over time. Therefore, the agent needs to take action to address it. For example, an agent with a low level of satisfaction in the \textit{belonging} need will not refill it unless it spends some leisure time with family and friends. However, as there are environmental constraints, if there are no agents in the leisure spots or they are at full capacity, the expected need will remain unsatisfied. Apart from positional restrictions, the model also contains financial limitations for certain spending-related actions. For example, agents with a low level of satisfaction regarding \textit{food} will be unable to go grocery shopping if their wealth attribute is lower than the price of food. In essence, the AABM is constrained by positional, financial and personal restrictions, related to the agents' personal profile and needs.

Mathematically, we denote the set of need categories as  $\mathcal{C}$ and the set of needs within each category $c \in \mathcal{C}$ as $ \mathcal{N}_c$. An \textit{importance function} maps every need category to the weight that the agent assigns to it, Imp : $\mathcal{C} \rightarrow[0,1]$. At time-step $t$, the need satisfaction level of need $n \in \mathcal{N}_c$ (for some $c \in \mathcal{C}$) is given by the need satisfaction level function $\mathrm{NSL}_t:\left\{\bigcup_{c \in \mathcal{C}} \mathcal{N}_c\right\} \rightarrow[0,1]$, which maps every need (across all categories) to its current degree of fulfilment. It can be written either as an iterative or exponential function
\begin{eqnarray}
    \mathrm{NSL}_{t}(n) = \gamma_{n,s} (n) \cdot \mathrm{NSL}_{t-1} (n)  \; \;\text{iterative or } \\   \mathrm{NSL}_n(t) = \gamma_{n, s}^{t} \cdot \mathrm{NSL}_n(0) \; \; \text{exponential} ,
    \label{eq: nsl}
\end{eqnarray}
where $\gamma_{n, s} \in [0, 1]$ is the decay rate for a need $n$ and an agent with status $s$. The status $s$ is a categorical variable indicating whether an agent is employed, unemployed, retired or homeless. The customization of the decay rates based on the nature of need and the agent's status $s$ allows us to create a differentiation between the personal circumstances of the agents. 

The iterative expression computes the current $\mathrm{NSL}_{t} (n)$ value by multiplying the decay with the previous $\mathrm{NSL}_{t-1} (n)$ value, while the exponential function is constituted by the $y$-intercept $\mathrm{NSL}_{n} (0)$, (the initial value of a particular need $n$), the decay rate and the current $\mathrm{NSL}_n(t)$ value. In order to include variability in the agent's lives, the initial values of the needs are sampled following a random normal distribution $\mathcal{N} (\mu_s, \sigma_s)$, with mean $\mu_s$ and standard deviation $\sigma_s$ dependant on status $s$.
\begin{figure*}[t]
    \centering
    \includegraphics[width=0.95\linewidth]{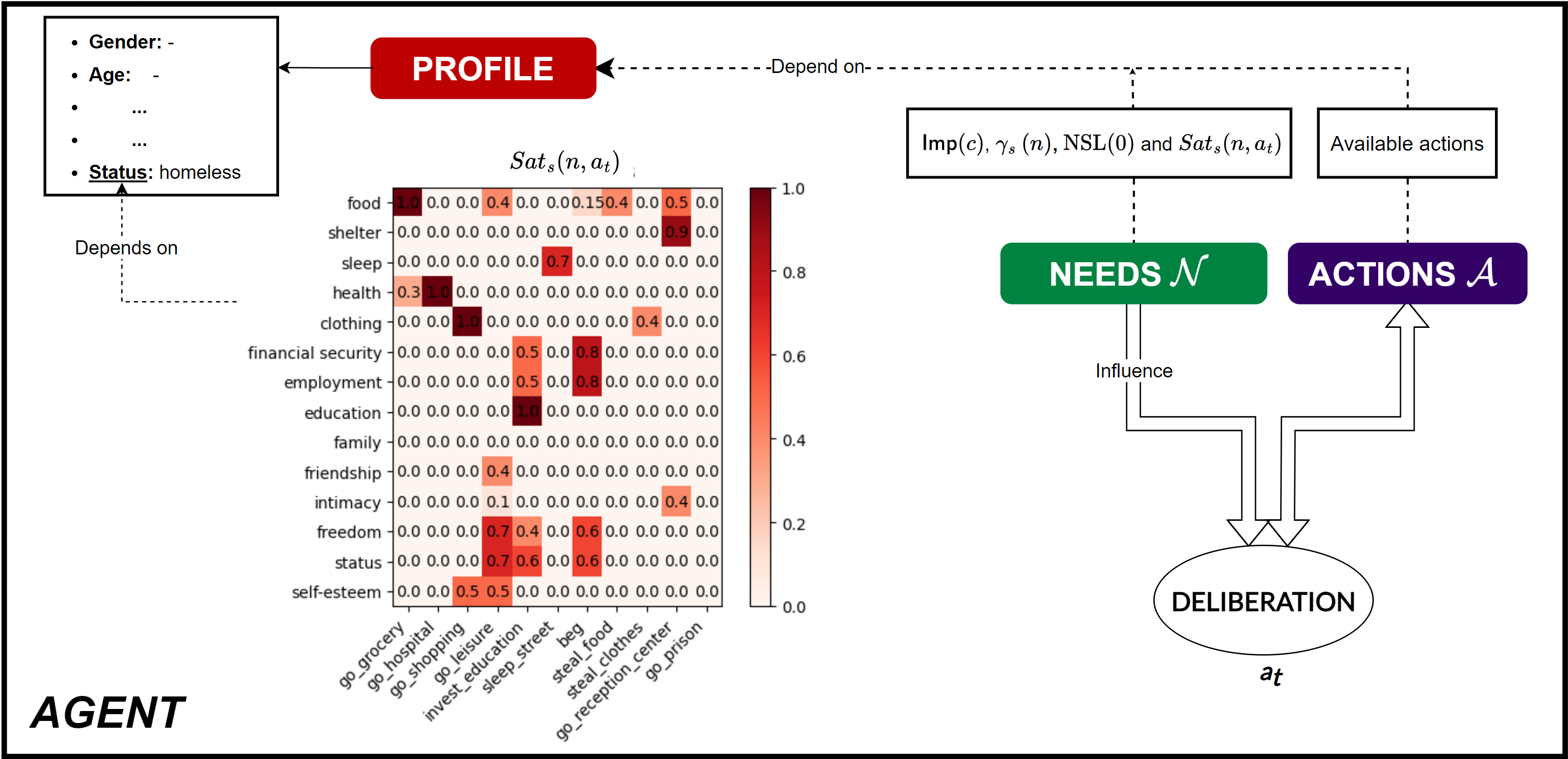}
    \caption{Scheme of the needs-based model implementation. The agent is defined by a profile which influences the deliberation of the action $a_t$ at timestep $t$. In turn, the numerical parameters and the available actions depend on the profile of the agent, particularly, on its status. The Expected Satisfaction matrix, $Sat_s (n, a_t)$, is represented in red for the homeless status.}
    \Description{Scheme showing the dependences of the needs, needs parameters and actions on the profiles. The Expected Satisfaction matrix is also shown. }
    \label{fig: scheme}
\end{figure*}
Conversely, the urgency of need $n$ at time step $t$ is defined as $\operatorname{Urg}_t(n)=1-\mathrm{NSL}_t(n)$, which implies that the urgency of the need will be higher when the need satisfaction level is low. Agents perform actions with the intention of refilling their most urgent needs, i.e. those with the highest depletion and following Maslow's pyramid. 

The set of actions available to an agent is denoted by $\mathcal{A}$ and is determined by the current status of the agent. For instance, only an employed agent can take action "\textit{go work}" or only a homeless agent can execute "\textit{go reception centre}". \footnote{A reception centre is a facility providing aid, shelter, and food to those in need.} To decide on which action to take next, agents consider the satisfaction they expect to draw from the execution of that action. The expected satisfaction function $Sat_s (n, a_t) :$ $\left\{\bigcup_{c \in \mathcal{C}} \mathcal{N}_c\right\} \times \mathcal{A} \rightarrow[0,1]$ captures these estimations by indicating the level of satisfaction the agent with status $s$ expects to get for need $n$ after executing action $a$. It is implemented as a matrix, where needs constitute the rows and actions form the columns, yielding dimensions $Dim(Sat_s) = n \times a$ for each agent status $s$. In Figure \ref{fig: scheme}, we illustrate this matrix for the homeless status. The matrix entries span from zero to one, with a maximum value indicating a strong correlation between the action and the corresponding need, and a minimum value indicating no effect of the action on the need. Intermediate values are assigned to actions that have a partial impact on the need, even if the relationship is not completely direct. For instance, the action-value pairs "\textit{go home}" and \textit{shelter} or "\textit{go grocery}" and \textit{food} have a maximum correlation, while pairs like "\textit{go home}" and \textit{intimacy} or "\textit{go leisure}" and \textit{food} exhibit a certain degree of correlation and are assigned intermediate values. Conversely, some actions do not correlate with certain needs, such as "\textit{go leisure}" and \textit{employment}. Additionally, there may be actions that do not satisfy any need, such as "\textit{go prison}". 
\begin{comment}
\begin{figure}[H]
    \centering
    \includegraphics[width=0.95\linewidth]{Images/homeless sat.png}
    \caption{Expected Satisfaction Level matrix ($Sat_{s}$) associated to the status $s$ corresponding to \textit{homeless}.}
    \label{fig: SAT}
\end{figure}
\end{comment}

\subsubsection{Deliberation} 

At each time step $t$, the agent deliberates about what is the best action $a_t$ to perform at the current time step, considering its available actions and its needs state. To decide, a score is computed for every action based on the level of satisfaction the agent is going to draw from executing it, across all of its needs weighted by the current urgency and the importance of their category. The agent selects the action with the largest score to perform. The formal expression of this deliberation function to determine the action $a_t$ is 
\begin{eqnarray}
    a_t & = & \underset{a \in \mathcal{A}}{\arg \max }\left[\sum_{c \in \mathcal{C}}\left(\sum_{n \in \mathcal{N}_c} \operatorname{Sat_s}(n, a) \cdot \operatorname{Urg}(n)\right) \cdot \operatorname{Imp}(c)\right].
\label{eq: decide action}
\end{eqnarray}

Immediately after determining the action, and for every need $n$, the need satisfaction level is updated according to the executed action and the $Sat(n, a_t)$ value associated with it. The updated need satisfaction level equation can be expressed as
\begin{eqnarray}
    \mathrm{NSL}_{t'} (n) = \mathrm{NSL}_{t} (n) + \alpha \cdot\operatorname{Sat_s} (n,a_t),
\label{eq: NSL}
\end{eqnarray}
where $\alpha$ is a parameter set to $0.5$ for all simulations, and $\mathrm{NSL}_{t'}$ is the value resulting from the addition of the $Sat_s (n,a_t)$ entry (dependent on the executed action at that step). When there is no correlation between the action and the need, $\operatorname{Sat_s} (n,a_t) = 0$ and $\mathrm{NSL}_{t}$ is equal to $\mathrm{NSL}_{t'}$. 

\begin{table*}[t]
	\caption{Policies (expressed as norms) implemented in the AABM, tagged as aporophobic (Apo) or non-aporophobic (Non-Apo). The description of the norms and a simplified codification (divided in precondition and post-condition statements) is included, as well as the legal reference of each policy. }
	\label{tab:normssum}
	\begin{tabular}{llp{5.05cm}p{3.6cm}p{3.5cm}r}\toprule
		\textit{Tag} & \textit{Norm (Id)} & \textit{Description}  & \textit{Precondition ($\phi_{Ag}$)}& \textit{Post-condition ($f$)} & \textit{Legal Ref.} \\ \midrule
		\multirow{3}{*}{Non-Apo} & (1) & 
Receive unemployment benefits if
you have fulfilled the required contributions when a month has gone by. & agx. status == ""unemployed""  and agx. time == 1 month &  agx. wealth += 700   & \cite{BOE_atur} \\%\cite{BOE_atur} \\ 
& (2) & 
Receive minimal vital income
when a month has gone by.  & agx. wealth \textless = 0 and agx. time == 1 month &   agx. wealth += 735  &  \cite{JefaturadelEstado2021} \\  
& (3) & Receive a dignified living space in
case of being homeless when a month has gone by.  & agx. status == ""homeless""  and agx. time == 1 month  & agx. home == new home and self. status == unemployed & \cite{ConsorciDHabitatgedeBarcelona2016}\\ 
\midrule
    \multirow{3}{*}{Apo} & (4) & 
Pay a fine when you sleep on the street
or commit a minor crime.  & agx. sleep street or agx. steal food and agx. wealth >= 0  &  agx. wealth -= 
500  & \cite{llei_multa_dormir} \\  
 &  (5) & 
In case you can not pay the fine
for a crime, the fine can be commuted
to days of imprisonment. & agx. sleep street or agx. steal food and agx. wealth \textless = 0  & agx. go prison and agx. depth += 500  & \cite{cod_penal} \\  
& (6) & 
Evicted from your home
in a bankruptcy situation when a month has gone by.  & agx. wealth \textless = 0  and agx. time == 1 month  &  agx. home == None  & \cite{BOE_eviction} \\  
\bottomrule
	\end{tabular}
\end{table*}

All the defined functions grant the agent the autonomy to choose the actions to be executed and derive their benefits. These decisions, directly affecting the agents' financial and personal situation, characterize the agent's behaviour, ultimately dictated by the weights, decays and expected satisfaction values. Thus, considering the importance of the numerical parameters associated with the behaviour of the agents, these must be fine-tuned. As they are, in turn, dependent on the personal status of the agent, we can accomplish this through a preliminary trial for each distinct agent status. Mainly, we examine and adjust the overall tendency of the wealth values, the frequency distribution of the executed actions and the evolution of the need satisfaction levels. Therefore, we are able to ensure that the weights, decays, and expected satisfaction values align with the model's temporality and the typical agents' behaviour.

\subsection{Regulatory Environment} \label{sec: regulatoryenv}
The third and final phase involves integrating a set of norms, representing the regulatory environment, into the simulation. The regulatory environment plays a central role in our model because it acts as an instance of aporophobia in our society. We aim to observe how these policies influence the final wealth distribution, providing evidence that discrimination against the poor has an impact on poverty levels. 

We make several assumptions concerning these policies. Firstly, we consider that all agents are fully aware of the rules and adhere to them perfectly. Secondly, it is assumed that the policies are flawlessly implemented, without any agents responsible for their enforcement. In other words, we do not model the potential autonomy of law-enforcing entities. Thirdly, we assume that the actions of the agents do not influence the selection of policies in any way, as there is no legislative process (e.g. voting) involved. These assumptions are made to focus on observing the effects of policies rather than the policy-making process itself, abstracting away any government bodies from the AABM \cite{AMPM}.

The different regulations are implemented as the verification of some precondition and the corresponding execution of the post-condition. The precondition and post-condition can be related to specific profile features of the agents or their available actions. Every norm is formalized as ($id$, $tag$, $\phi_{Ag}$) $\rightarrow$ $f$, where $id$ is the identifier of the norm, $tag$ is its tag (Apo or NonApo) and $\phi_{Ag}$ is a precondition that is checked in the context of an agent (i.e. to test whether the rule should apply to the agent). The post-condition $f$ is modelled as a callable function that ought to be applied to the agent, provided that the precondition holds true. 

\section{Use case} \label{sec: use case}
As a first case study of the simulation, we choose 4 districts within the city of Barcelona: Sarrià-Sant Gervasi, Gràcia, Les Corts and Eixample. The data is mainly extracted from OpenData Ajuntament de Barcelona databases \cite{Open_data}. Other sources provided by NGOs include the population that is not represented in the city council’s databases, such as homeless or immigrant persons without an address nor an up-to-date administrative situation \cite{arrels}. The demographic data is used to define the agents' profile by including the distribution of different attributes already mentioned (e.g. age, gender, rent, income) in each district. All these demographic and socio-economic variables are transformed into probability distributions, from which agents' profiles are sampled, ensuring they are representative of real-world demographics. 

\begin{figure*}[t]
    \centering
    \begin{minipage}[c]{0.285\linewidth}
        \includegraphics[width=\linewidth,valign=c]{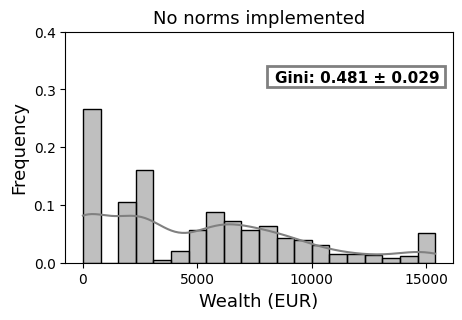}
    \end{minipage}
    \qquad
    \begin{minipage}[c]{0.67\linewidth}
        \includegraphics[width=\linewidth]{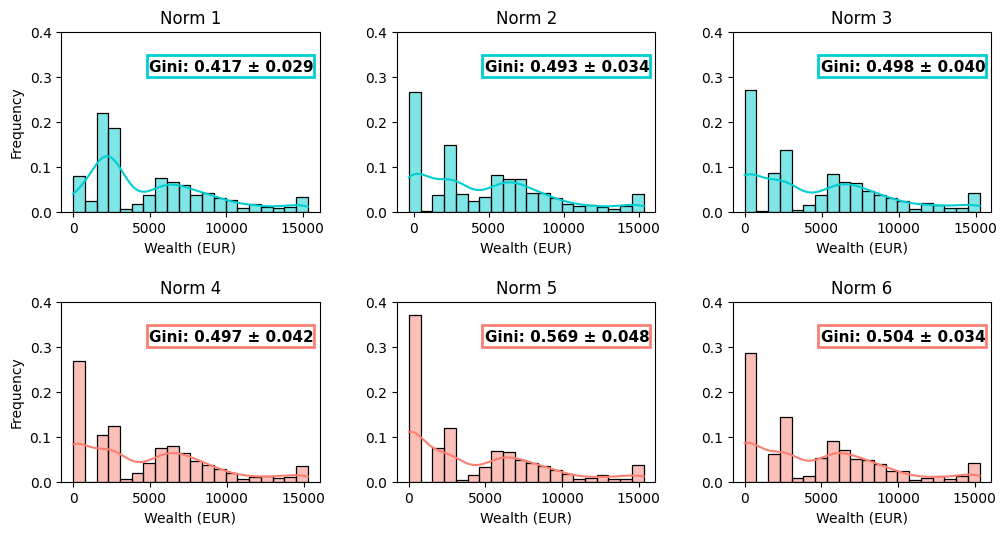}
    \end{minipage}
    \caption{Results extracted from the AABM's proof of concept for the examination of the individual effect of norms. Wealth distribution for the scenario with no norms implemented is represented in grey, while the distributions associated with Apo-tagged and Non-Apo tagged norms are depicted in pink and blue respectively, with the corresponding Gini coefficient. } %, with the corresponding Gini coefficient and the density probability line smoothing the distribution.
    \Description{Results extracted from the AABM's proof of concept as final wealth distributions. The figure comprises 8 plots arranged in a 2x4 grid. At the left top and bottom, we present the model execution with No norms implemented (baseline simulation), a grey-coloured wealth distribution with an associated Gini coefficient equal to ($0.481 \pm 0.029$). Right next to it, the remaining 6 plots correspond to the wealth distributions outputted from the execution of the model being affected by each of the norms individually. The Gini coefficients associated with them are, for each of the Norms respectively: Norm 1 ($0.417 \pm 0.029$), Norm 2 ($0.493 \pm 0.034$), Norm 3 ($0.498 \pm 0.040$), Norm 4 ($0.497 \pm 0.042$), Norm 5 ($0.569 \pm 0.048$) and Norm 6 ($0.504 \pm 0.034$). }
    \label{fig: resultsind}
\end{figure*}

In our simulation use case, we perform some adaptations to simplify the needs-based model methodology explained in section \ref{sec: needsmodel}. These adjustments regard two key aspects:  (1) Maslow's hierarchy and (2) the decision-making time frame. In the first place, we disbar the growth needs (self-actualization category of Maslow's hierarchy of needs in Fig. \ref{fig: maslow}) from the simulation. Maslow's theory of needs states that while every individual possesses the capability and aspiration to progress up the hierarchy towards self-actualization, this advancement is often impeded by the failure to fulfil lower-level needs \cite{maslow1943theory}. As our main focus is on modelling behaviour related to basic needs, we only consider deficiency needs, basic and psychological needs constituting the lower level of the pyramid. Secondly, the autonomous decision-making feature is not operational at every time step of the model. The decision-making process starts at $5$ pm for all agents. For the rest of the agents' day, certain actions are fixed by the time of the model to ensure the duties of the agents are executed (regardless of their needs). Essentially, employed agents are forced to go to work for a set amount of time, and all agents are forced to sleep at night. 

As an initial prototype, a set of six norms, denoted as $\mathcal{N_o}$, is codified and tested in the model. Table \ref{tab:normssum} introduces the norms implemented, with a simple explanation of each one of them and their legal reference, mostly from the official Spanish government gazette, Boletín Oficial del Estado (BOE) \cite{BOE}. The codification of the norms appears as two simplified precondition and post-condition coding statements. The aporophobic or non-aporophobic tag of each norm has been obtained by collaborating closely with legal NGOs specialized in poverty mitigation and homelessness \cite{arrels}. 

\section{Results} \label{sec: results}
To comprehensively examine the impact of the norms proposed in Table \ref{tab:normssum}, we run several executions of our AABM for all the combinations (subsets) of norms, presented in section \ref{sec: regulatoryenv}. This corresponds to the power set of the norms, which is denoted as $\mathcal{P (N_o)_{}}$, and equals $ 2^6 = 64$. It includes all subsets in $\mathcal{N_o}$, including the empty set (no norms are enforced at all). Then, our results will map the elements in $\mathcal{P (N_o)}$ to an indicator of wealth inequality.

Given the randomness inherent to our model (e.g. in initializing agent wealth and other attributes) it is necessary to generate a sample of runs ($\tau = 10$) for each subset of norms, with a subsequent standard deviation associated with the results. We choose to execute the model on the IIIA's high-performance computing infrastructure \footnote{\url{https://iiia.csic.es/en-us/research/ars-magna/}}, using \texttt{mesa.batch\_run} to parallelize the executions. All simulations are executed for $T = 2880$ timesteps (the equivalent of 4 months), with $N = 100$ agents.

Every rule configuration is assessed in terms of the final wealth distribution, collected at the end of every simulation run. The final wealth distribution comprises the sum of all the simulations performed with a particular combination of norms, meaning that the total count of agents is always $N \cdot \tau = 1000$, normalized to $1$ in the $y$-axis. This frequency axis refers to the proportion of agents with final wealth falling within the corresponding value. The distribution is smoothed with a probability density function, a line associated with the likelihood of the variable falling within a certain range. All the results obtained are compared with the baseline case (the simulation with no implemented norms, depicted in grey in all of the cases). For the purpose of the analysis, the column corresponding to the proportion of agents ending up with the lowest wealth (left bar of the graphs) will be denoted as \textit{bankruptcy bar}.

We measure the impact of the regulatory environment on wealth distributions by using the Gini coefficient, an economic measure that quantifies the degree of wealth inequality within a social group. It ranges from zero to one, with a null Gini coefficient reflecting perfect equality (i.e. all members in a community own identical shares of wealth), and a maximum Gini coefficient reflecting maximum inequality (i.e. all wealth is concentrated in a few individuals)\cite{gini}. 
\begin{figure*}[t]
     \centering
     \begin{subfigure}{0.33\textwidth}
         \centering
         \includegraphics[width=\textwidth]{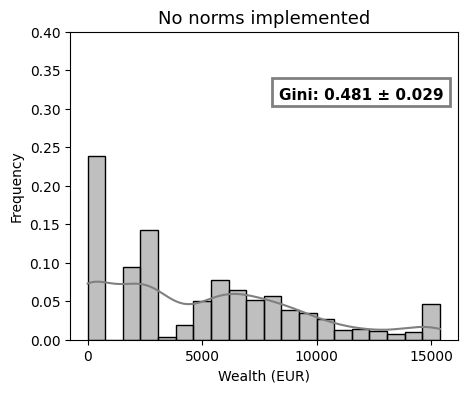}
         \caption{}
         \label{fig: resultsa}
     \end{subfigure}
     \begin{subfigure}{0.33\textwidth}
         \centering
         \includegraphics[width=\textwidth]{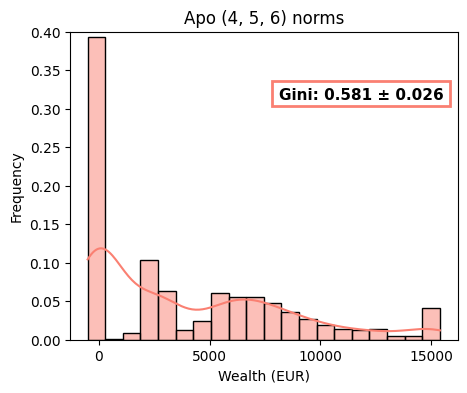}
         \caption{}
         \label{fig: resultsb}
     \end{subfigure}
      \begin{subfigure}{0.33
      \textwidth}
         \centering
         \includegraphics[width=\textwidth]{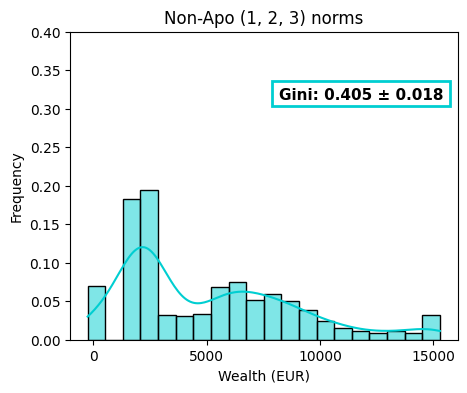}
         \caption{}
         \label{fig: resultsc}
     \end{subfigure}
        \caption{Results extracted from the AABM's proof of concept for the examination of the collective effect of norms. (a) Wealth distribution without implemented norms depicted in grey. (b) Wealth distributions implementing the subset of norms tagged as Apo (\textit{Norms 4, 5, and 6}) depicted in colour pink. (c) Wealth distributions implementing the subset of norms tagged as Non-Apo (\textit{Norms 1, 2, and 3}) depicted in colour blue.  }
        \label{fig: results}
        \Description{Results extracted from the AABM's proof of concept as final wealth distributions. The figure comprises 3 subplots. At the left, exactly as in the previous Figure, we present the model execution with No norms implemented (baseline simulation), a grey-coloured wealth distribution with an associated Gini coefficient equal to ($0.481 \pm 0.029$). Right next to it, the wealth distributions corresponding to the execution of the model are affected by the Apo and Non-Apo subsets of the norms collectively. The Gini coefficients associated with the Apo and Non-Apo subsets are $(0.581 \pm 0.026)$ and ($0.405 \pm 0.018$), respectively.}
\end{figure*}
Since there are numerous subsets of norms, we focus on studying the effects of some subsets in particular: every norm individually (section \ref{sec: resultsind}), the aporophobic and non-aporophobic subsets of norms, Apo (\textit{Norms 4, 5, and 6}) and Non-Apo (\textit{Norms 1, 2, and 3}) respectively (section \ref{sec: resultscoll}), and all subsets together (section \ref{sec: resultsgini}). 

\subsection{Individual Effect of the Norms} \label{sec: resultsind}
The effect of the norms can be analyzed in isolation. Fig. \ref{fig: resultsind} reveals that most norms in scope individually increase the Gini coefficient compared to the baseline simulation without implemented norms. Only \textit{Norm 1} (tagged as Non-Apo) decreases the Gini coefficient. \textit{Norm 2, 3, 4 and 6} have a similar effect on the Gini, showing an increment of $3.5 \%$ approximately ($4.7 \%$ in the case of \textit{Norm 6}). On the other hand, \textit{Norm 5} significantly raises the coefficient by $18.3\%$.

One can see that the \textit{bankruptcy bar} is the lowest for \textit{Norm 1}, indicating that it is the norm that most effectively reduces the proportion of agents ending up in a bankruptcy situation. This is reasonable because \textit{Norm 1} affects a significant amount of agents, susceptible to ending up broke, by providing them with unemployment benefits. 

The ineffectiveness of the rest of the Non-Apo tagged norms in decreasing the Gini can be attributed to the wide-ranging and low-ranging effects of the \textit{Norms 2} and {3} on the agents. \textit{Norm 3} affects all agents equally in terms of minimal vital income, while \textit{Norm 2} only affects homeless agents by providing them with a dignified living space. Therefore, implementing \textit{Norm 3} for all agents does not significantly alter the overall distribution, and implementing \textit{Norm 2} for a small portion of the total number of agents has a minimal effect on the wealth distribution. 

This can be further examined by analyzing the proportion of agents affected within the \textit{bankruptcy bar} and their respective statuses. The baseline simulation leads 23.84\% of the agents to a bankruptcy situation, 71.1\% of them unemployed and 28.3\% homeless. \textit{Norm 1} improves meaningfully this situation, with only 7,9\% of the population, all with homeless status assigned, falling within the bankruptcy bar. On the other hand, \textit{Norm 2} and \textit{Norm 3} result in a very similar proportion of financially broke agents (26.6\% and 27.0\%), but with different statuses associated ($68.5\%$ unemployed and $31.5\%$ homeless in the first one, and $99.1\%$ unemployed and $0.9\%$ employed in the second one). This indicates that, although the bankruptcy bar seems to be decreased similarly, the final state of the isolated scenarios differs by the status of the agents. \textit{Norm 2} (minimal vital income scenario) is dealing with homeless and unemployed agents, whereas \textit{Norm 3} (dignified living space scenario) is not contemplating homeless agents anymore, since they are already settled. Inside the Non-Apo tagged norms, \textit{Norm 5} (paying a sanction or commuting it with days of imprisonment) leads to the highest bankruptcy bar (37\% of the agents, tagged with homeless status).

%Inside the Non-Apo tagged norms, \textit{Norm 5} (related to paying a sanction and going to prison) leads to the highest bankruptcy bar (37\% of the agents, tagged with homeless status). Since this norm leads agents in bankruptcy situations to transition their status to homeless, it is reasonable to obtain only homeless agents within the bar.

\subsection{Collective Effect of Norms: Apo and Non-Apo subsets} \label{sec: resultscoll}

It is interesting to examine the joint effect of the subset of Norms tagged as Apo and Non-Apo. The results for the Apo and Non-Apo subsets appear in Figure \ref{fig: resultsb} and \ref{fig: resultsc}. By comparing them with Figure \ref{fig: resultsa}, one can see that the Non-Apo subset of norms decreases the Gini by 18.8\% (leads to a more egalitarian distribution of wealth), while the Apo subset increases it by 20.8\% (leads to a less egalitarian distribution of wealth). 

It is worth noticing that the bankruptcy bar is significantly lower for the Non-Apo combination, indicating the effectiveness of Non-Apo policies in preventing bankruptcy situations. For Non-Apo norms, the analysis reveals 6.96\% of the agents experiencing bankruptcy (93\% unemployed and 6\% employed). In contrast, Apo norms result in $39.3\%$ of homeless agents facing bankruptcy. Non-Apo norms policies directly provide shelter to homeless agents, thereby categorizing them as unemployed, providing them with unemployment benefits and offering educational opportunities. Conversely, Apo norms contribute to an increase in homelessness and bankruptcy. 
\begin{figure}
    \centering
    \begin{subfigure}{\linewidth}
        \includegraphics[width=0.97\linewidth]{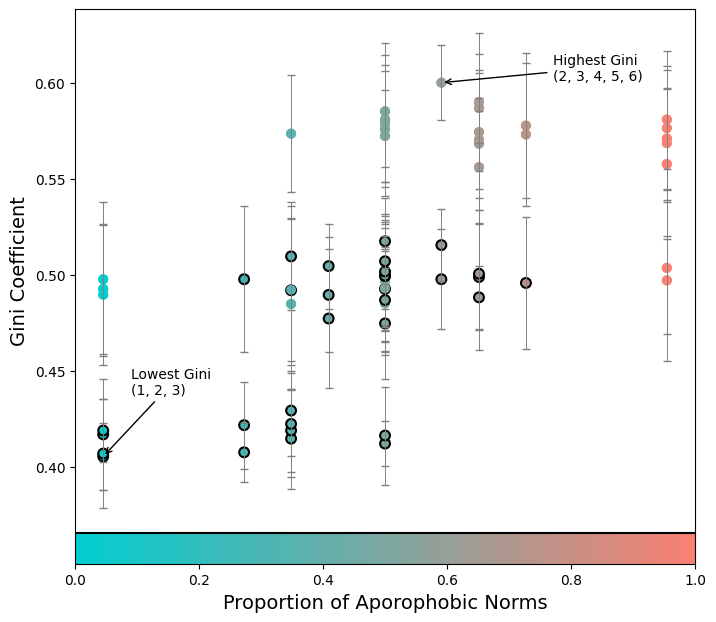}
        \caption{}
        \label{fig: gini}
    \end{subfigure}

    \begin{subfigure}{\linewidth}
        \includegraphics[width=0.96\linewidth]{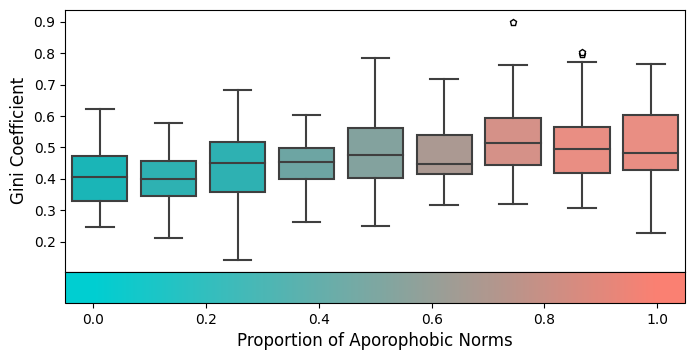}
        \caption{}
        \label{fig:boxplot}
    \end{subfigure}
    \caption{Results extracted from the AABM's proof of concept across all combinations of norms. (a) Gini coefficients as a function of the proportion of aporophobic norms in each norm combination executed. Maximum aporophobia proportion is depicted in pink, while minimum aporophobia is represented in blue. Lowest and Highest Gini computed are indicated, as well as their associated combination of norms. All the Gini coefficients obtained with a combination containing \textit{Norm 1} are circled in black. (b) Gini coefficient variation illustrated with boxplots for every norm combination associated with a certain proportion of aporophobic norms.}
    \Description{ (a) Scatter plot of all the possible Gini coefficients as a function of the Proportion of Aporophobic Norms in the combination. A color bar indicates the color associated with minimum Aporophobia (blue) and maximum aporophobia (salmon). (b) Boxplot of the individual variation of the Gini Coeficients for the 9 different values of Proportion of Aporphobic norms in the combination. Some outliers can be detected. }
    \label{fig:gini_combined}
\end{figure}

\subsection{Influence of Aporophobia in the Gini Coefficients} \label{sec: resultsgini}

As a final observation, we analyze the ensemble of all norm combinations and their impact on the computed Gini coefficient. Figure \ref{fig: gini} illustrates the Gini coefficients across all different combinations of norms, as a function of the proportion of Aporophobic-tagged norms. Generally, a higher proportion of Aporophobic norms (maximum value of 1) corresponds to higher Gini coefficients, located on the right side of the plot. Conversely, Non-Aporophobic norms lead to lower values, located on the left side.

This type of plot reveals the combination of norms that leads to the highest and lowest Gini coefficient, indicated in the graph, allowing the examination of intermediate values too. The lowest Gini obtained is ($0.405 \pm 0.018$), achieved by implementing the combination of \textit{Norms (1, 2 and 3)}, which corresponds to the Non-Apo subset (nule proportion of Aporophobic tagged norms). However, the highest Gini coefficient obtained is ($0.600 \pm 0.020$), which is the result of the combination of \textit{Norms (2, 3, 4, 5 and 6)}, with $60\%$ of Aporophobic-tagged norms in it. Furthermore, we analyze isolatedly the Gini coefficients obtained as a result of a combination of norms containing \textit{Norm 1} in it (circled in black in Figure \ref{fig: gini}). The results evidence that \textit{Norm 1} is contributing significantly to the most egalitarian outcome because most of the lowest and intermediate Gini values outcomes incorporate this norm.

The standard deviation of the Gini coefficients, obtained for the macroeconomic distributions of each combination, has been included to address the results’ variation. However, the individual variation of the $10$ different wealth distributions is not visualized in the plots. We can examine this variation by looking at the box plot, associated with the distribution of Gini coefficients, for each configuration of norms in Fig. \ref{fig:boxplot}. Notice that the interquartile range coincides approximately with the Gini values obtained above.

\section{Conclusions, limitations and future work} \label{sec: concl}
In this paper, we present the AABM, a novel multi-agent autonomous decision-making simulation that assesses the potential impact of policies in terms of wealth inequality within a specific population. In this initial version, the policies in scope focus on addressing homelessness and poverty in 4 districts of Barcelona. The model is built on the needs-based model by Dignum et. al \cite{dignum2020analysing}. By initializing the agents based on real demographic data, we accomplish a meaningful representation of the citizens residing in the city districts. 

We provide evidence of the correlation between norms tagged as aporophobic (considered discriminatory against the poor by non-profits) and wealth inequality, which affects poverty reduction rates. The results fit our expectations: non-aporophobic policies seem to lead to a more egalitarian environment. These insights aim to contribute to the number one UN Sustainable Development Goal, a global call for action to mitigate poverty, by informing a new generation of poverty reduction policies that tackle poverty not only by acting on the redistribution of wealth, but also by mitigating discrimination against the poor at an institutional level. %(in the regulatory system).

This model opens a wide range of possibilities to explore in terms of policy assessment before implementation. Nevertheless, it currently stands as a proof of concept with limitations that need to be addressed in future work. First of all, in order to make the AABM more robust, the agent population needs to be completely representative of the real population. To achieve this, the model will incorporate the demographic data in terms of a validated synthetic population. This synthetic population will be expandable to include additional attributes, thereby enriching the contextual behaviour of agents in each region 
 \cite{synthpop}. Secondly, we will consider a wider range of policies that might influence poverty and inequality, such as regulations regarding education. Thirdly, we aim to obtain outputs from the AABM in terms of levels of homelessness, evictions, education or other indicators, which reflect the multidimensionality of poverty and can be more meaningful for non-profits and government officials. The three ways forward are currently being developed to explore the potential of the AABM as a regulatory tool. The enhancement of the model will allow us to offer rigorous comparative studies for similar policies in different urban environments. % Additionally, we aim to validate the model through historical data on Gini coefficients \cite{gini_barcelona}.  

\begin{acks}
This research has been supported by the EU-funded VALAWAI (\#~101070930), the Spanish-funded VAE (\#~TED2021-131295B-C31) and the Rhymas (\#~PID2020-113594RB-100) projects. Thanks to Beatriz Fernández Gensana, Head of the Juridic Team in \textit{Arrels Fundació} (\url{www.arrelsfundacio.org}), who help us as a legal counsellor to select, manage and understand the policies implemented for the city of Barcelona.  
\end{acks}

%%%%%%%%%%%%%%%%%%%%%%%%%%%%%%%%%%%%%%%%%%%%%%%%%%%%%%%%%%%%%%%%%%%%%%%%

\bibliographystyle{ACM-Reference-Format} 
\bibliography{AAMAS_2024}

%%%%%%%%%%%%%%%%%%%%%%%%%%%%%%%%%%%%%%%%%%%%%%%%%%%%%%%%%%%%%%%%%%%%%%%%

\end{document}